# Predictability of localized plasmonic responses in nanoparticle assemblies

Kevin M. Roccapriore[1], Maxim Ziatdinov[1,2], Shin Hum Cho[3,4], Jordan A. Hachtel[1], and Sergei V. Kalinin[1,2]

[1] Center for Nanophase Materials Sciences, Oak Ridge National Laboratory, Oak Ridge, TN 37831

[2] Computer and Computational Sciences Division, Oak Ridge National Laboratory, Oak Ridge, TN 37831

[3] McKetta Department of Chemical Engineering, The University of Texas at Austin, Austin, Texas 78712, United States

[4] Samsung Electronics, Samsung Semiconductor R&D, Hwaseong, Gyeonggi-do 18448, Republic of Korea

Design of nanoscale structures with desired nanophotonic properties are key tasks for nanooptics and nanophotonics. Here, the correlative relationship between local nanoparticle geometries and their plasmonic responses is established using encoder-decoder neural networks. In the *im2spec* network, the correlative relationship between local particle geometries and local spectra is established via encoding the observed geometries to a small number of latent variables and subsequently decoding into plasmonic spectra; in the *spec2im* network, the relationship is reversed. Surprisingly, these reduced descriptions allow high-veracity predictions of the local responses based on geometries for fixed compositions and chemical states of the surface. The analysis of the latent space distributions and the corresponding decoded and closest (in latent space) encoded images yields insight into the generative mechanisms of plasmonic interactions in the nanoparticle arrays. Ultimately, this approach creates a path toward determining configurations that can yield the spectrum closest to the desired one, paving the way for stochastic design of nanoplasmonic structures.

---

[1] Corresponding author, roccapriorkm@ornl.gov
[2] Corresponding author, sergei2@ornl.gov

Localized surface plasmon resonances (LSPRs) are collective oscillations of the free charge in nanostructures that concentrate electromagnetic energy and enable the enhancement of wave-matter interaction as well as the manipulation of light at nanometer length-scales. Moreover, plasmon resonances are highly influenced by sample geometry and local dielectric environment, which enables the application of biological and chemical sensing[1]. State-of-the-art nanoscale synthesis is required to exploit the characteristics of plasmon resonances[2–6] which creates the opportunity to rationally design extremely complex systems with a plasmonic response tailored to accentuate specific phenomena such as engineered electric permittivities[7] and even cancer detection and treatment[8]. Finally, the need for light sources localized well below the wavelength of light for quantum applications and optical circuit elements[9,10] has given rise to extensive research efforts in this direction.

For highly ordered systems, such as precise lithographically patterned arrays, various effective structure-property relationships can be established through macroscopic experiments and simulations. For example, specially engineered arrays can be designed to produce specific effective properties not found in nature, e.g., negative refractive index metamaterials[11,12]. However, in these cases, the detection volume significantly exceeds the characteristic plasmon size. It follows that for disordered systems containing multiple local geometries and morphologies, macroscopic techniques can only sample the ensemble averages and effective responses, and therefore any localized behavior is missed. Furthermore, the randomness of the disorder necessitates extensive sampling of the various inhomogeneities to truly ascertain an accurate representation of the true nanoscale response and makes direct simulation of all possible types of disorder impractical.

The need to access local effects in plasmonic systems has driven significant interest in nanoscale spectroscopy techniques, one of the most powerful of which is electron energy-loss spectroscopy (EELS) in the scanning transmission electron microscope (STEM). Here, the electron beam acts as a white light source in which the high-energy electrons couple to the plasmonic material, which manifests as a loss of the electron energy at the available plasmon modes[13], yielding detectable peaks in the low-loss region of EEL spectrum. As the electron beam can be about the diameter of a single atom, this makes the electron microscope uniquely suited to study the nanoscale spatial behavior of plasmon excitations[14–16]. However, the small probe of STEM means that only a small field-of-view can be sampled efficiently per STEM-EELS experiment, preventing high throughput analysis of systems with large random variation in heterogeneity. As an alternative to direct spectral analysis, if structure-property relationships can be determined unambiguously, one could use the structure as determined through imaging analysis to determine the spectral response. This would be highly beneficial since dwell times in spectroscopy are generally in the hundreds (if not thousands) of milliseconds, while efficient structural imaging can be achieved with dwell times in the microsecond and even nanosecond regimes[17].

High-veracity prediction, however, of the local structure-property relationships is limited. Even in well-defined systems, calculations of non-local plasmonic responses are hindered by the computational complexity of predictive theory. In real systems, the presence of surface layers and adsorbates can result in a large number of poorly understood and weakly controlled variables. In fact, the very factors that enabled high sensitivity of nanoplasmonic structures to external stimuli severely complicate predictions of these behaviors. Moreover, the non-local collective response observed in many-particle plasmonic systems conceals physical mechanisms even further. This in turn brings an issue whether the mechanisms guiding the emergence of the plasmonic response can be understood and controlled, both in terms of the fundamental generative mechanisms and predictive models that can establish whether structures with required properties can be created. Thus, there exists an inherent need for a data-driven predictive methodology to empirically establish structure-property relationships in real complex plasmonic systems and enable rigorous high-throughput analysis.

Here, we explore a machine-learning (ML) approach for the exploration of nanoplasmonic structures based on developing parsimonious correlative laws between the local structure and plasmon response. This approach establishes the relationship between observed EEL spectra and semi-local particle geometries within a given materials system by using an autoencoder (AE) network and can be used to predict plasmonic responses in similar systems. The inverse problem, in which particle geometries are predicted from the observed spectra, is considerably more ill-defined. We propose that the areas where the correlative relationships between the observed particle geometries and spectra have the strongest deviations can serve as indicators for the manifestation of potentially new physical phenomena.

Here, we examine self-assembled monolayers of fluorine and tin doped indium oxide (FT:IO) nanocrystal arrays[18]. The geometry and the plasmonic response can be tuned by changing the doping concentration. As a result, the FT:IO particles can either possess either a cubic or spheroidal geometry with diameters between 10 and 20 nm, as well as possessing native plasmon frequencies ranging from the near- to mid-infrared (IR). Critically, while the self-assembled structures are nominally periodic, the colloidal synthesis process introduces a high-degree of localized disorder in the particle size and shape variation, missing particles (defects), and holes, cracks, and edges in the self-assembled films. As a result, understanding the nanoscale response of the system requires the nanoscale resolution of a technique like STEM-EELS to account for this heterogeneity. To feasibly detect the IR plasmonic response, the electron beam is passed through a pre-specimen monochromator, which reduces the elastic scattering background in the IR and improves energy resolution[19]. In these measurements, we utilize an energy resolution of ~40 meV, which allows us to observe the plasmonic excitations without significant instrumental broadening.

The hyperspectral EELS datasets used for these analyses are called spectrum images (SIs), which are obtained by rastering the beam through a region-of-interest and recording a full EEL spectrum at each probe position, resulting in a three-dimensional dataset with two-spatial dimensions and one spectral dimension. An SI of a heterogeneous FT:IO array is shown in Figure 1. The dark-field image of the array is shown in Fig. 1a, which shows that while the array is mostly regular, a hole (or defect) in the array as well as the edge of the film are present, both providing strong aspects of heterogeneity in the film. We show spectra from four selected regions in the SI (R1-the center of a particle, R2-the gap between particles, R3-the hole in the array, and R4-the area outside the array), and plot the averaged spectra at those locations in Fig. 1b. There are three prominent features in the EELS response: a dominant peak at ~500 meV, a subpeak at ~750 meV, and a small peak at ~900 meV, however we note that both the relative intensities as well as the frequencies of these peaks change significantly from position to position. These localized changes demonstrate that heterogeneity provides a localized impact on the plasmonic structure.

To gain further insight and explore the plasmonic behavior in the nanoparticle array in a less biased approach, the dimensionality of the 3D EELS data cube can be reduced used using linear unmixing methods. These methods in general decompose the hyperspectral dataset into components with a spatial abundance map and a spectral endmember, such that the linear combination of each component's endmembers weighted by their abundance reproduces the original dataset. There are numerous such unmixing techniques, but one method that tends to work well for EELS datasets is non-negative matrix factorization (NMF)[20]. The NMF decomposition is chosen due to the non-negativity constraint, which generally results in more physical spectral and spatial components for a counting spectroscopy such as EELS. A simplistic 4-component NMF deconvolution is shown in Fig. 1c-j, with the abundance maps shown in (c-f) and the corresponding spectral endmembers in (g-j).

While the spatial abundance maps very nicely highlight the different regions of heterogeneity in the array (particles, gaps, hole, edge/outside), the corresponding endmembers clearly have multiple plasmonic peaks per component, indicating that we have not fully separated the different mechanisms in the plasmonic response – in other words, the electron beam excites all plasmon modes, some of which unavoidably occur in the same location in space. This result follows naturally for heterogeneous quasi-periodic structures such as the FT:IO arrays. The periodicity and intra-particle coupling induce an overarching response in the array, thus the primary influence of heterogeneity is to modify this overarching plasmonic behavior locally as opposed to providing a distinct new one. This effect is replicated in the NMF decomposition, where component 1 captures the overall dominant response of the structure, while the higher order components capture the local modifications to that response due to the various structural features. Another important aspect captured by the NMF decomposition in Fig. 1 is the different degrees of localization. For instance, the first two components are highly delocalized extending well into the hole in the array and for tens of nm outside of the array, meanwhile the other two components correspond to signal that is highly localized to the individual nanoparticles themselves and do not extend more than a few nm outside of the edge of the cube. As a result, in order to accurately represent the response of the system, the predictive ability of any ML network must be able to reproduce the frequencies, intensities, and localization of the complex plasmonic response in the FT:IO arrays.

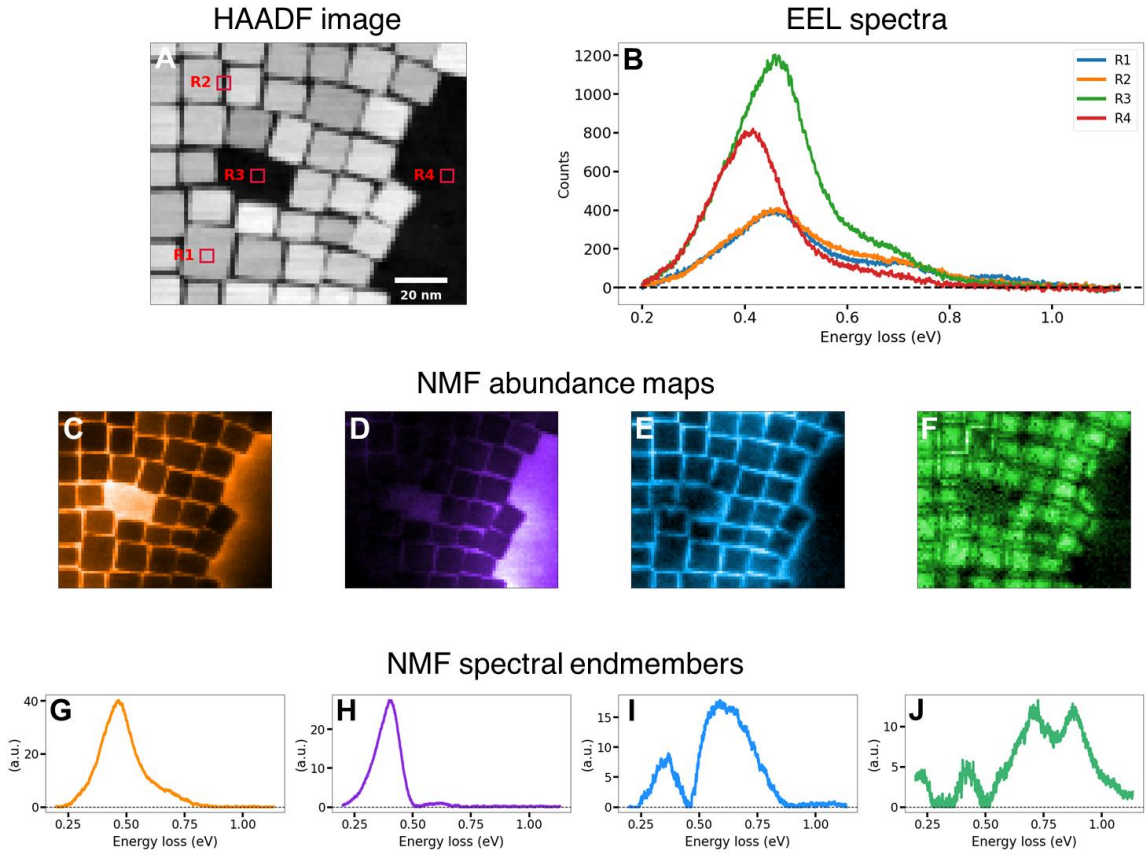

**Figure 1**. (a) HAADF-STEM image of nanoparticle array. (b) EEL spectra from selected locations (denoted by squares in (a)). (c-j) First four NMF components, (c-f) component abundance maps (g-j) component spectral endmembers.

The number of particles involved in complex heterogeneous systems, such as the ones shown in Fig. 1, is too high for effective direct calculation, necessitating an empirical data-driven approach. In our approach, we utilize deep learning networks trained only on the experimentally acquired data to empirically derive the relationship between local structure and local plasmonic response without theoretical predictions. To establish local structure-property relationships, we train an encoder-decoder network to extract a set of latent variables that connects spatial descriptors to spectral descriptors, where the spatial descriptors are sub-images of the structure surrounding a specific pixel in the SI, and the spectral descriptors are the EEL spectra from the SI at that specific pixel. We create networks to establish the correlative relationship between the spatial descriptors and spectral descriptors, ultimately in order to predict the spectral response given a spatial predictor, as well as the inverse prediction. These networks are called *Im2spec* and *Spec2im*, respectively. In both cases, the analysis workflow includes (i) analysis of latent space distributions, (ii) forward prediction from the latent space, comparison with the ground truth, and visualization of the error maps, and (iii) back-projection from the latent space (to the closest experimentally observed point).

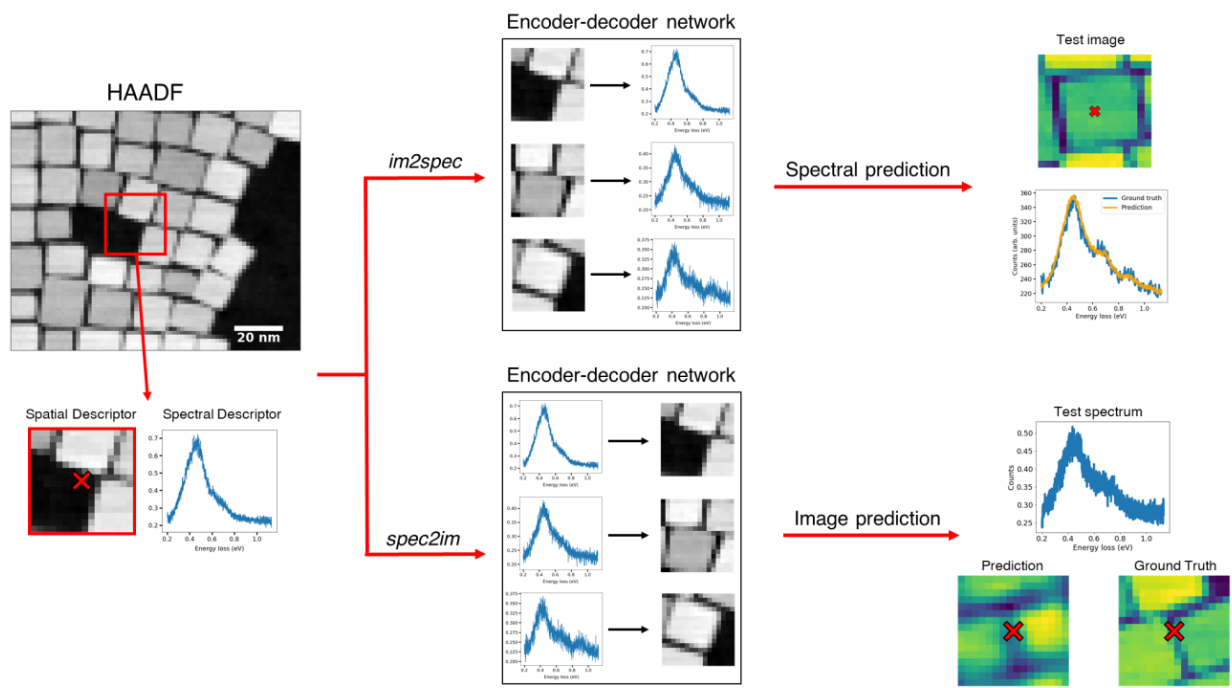

**Figure 2.** Prediction workflow using autoencoder architecture. Spectrum image is separated into small windows with a corresponding single associated spectrum from the center of the window. These pairs are sent to one of the encoder-decoder networks, where a portion of the pairs are used as training sets, and the remaining are used for testing. Prediction of the spectrum or image is performed and compared to the ground truth representation. The simultaneously acquired and therefore registered HAADF signal is used as the structural data to indicate sample geometry.

The overall workflow for the proposed approach is illustrated in Figure 2. The SI is analyzed to yield a feature/target set made from the local sub-images centered at specific locations and EEL spectra corresponding to the center of the sub-image. In the *im2spec* network[21], the images are the features and the spectra are the targets. In the *spec2im* network, the spectra are the features and images are the targets. Both *im2spec* and *spec2im* networks are based on an encoder-decoder architecture, i.e., the feature set is compressed through the set of the classical convolutional layers (2D for image data, 1D for spectral data) to a latent layer and subsequently "decompressed" to the target set via spatial pyramids of dilated convolutions. This approach both establishes the correlative relationship between the feature and target sets and allows exploration of the variability of the observed behaviors via low-dimensional representations in the latent space of the network filtering out the non-essential details. The *im2spec* and *spec2im* networks are implemented here via the PyTorch deep learning library. The schematic of the *im2spec* network is shown with more details in Figure S2, with specific methodologies explained in the **Methods** section.

Since the particles are colloidally synthesized, we expect them to be chemically identical, hence the primary differences between particles will be structural (i.e. size, shape, faceting, etc.). This suggests there should be a well-defined relationship between the local configurations of the particles and the associated plasmonic response. In general, the details for larger separations from the chosen pixel location become progressively less important, however, it should be noted that some delocalized responses can be relatively strong and there may be appreciable spectral strength even far from the chosen location (e.g., outside field of view), hence this must be taken into account. Even if such a relationship does exist, it is not necessarily guaranteed that it is observable, since the observations can also be affected by the latent variables (e.g., different contaminations or surface states for different particles and variations in particle morphologies), observational biases (configuration of imaging system and its variability across the image plane), and noise limits. Nevertheless, the exploratory data analysis in Fig. 1 shows that the abundance maps in the NMF components clearly depicts the presence of different structural features: particles, edges and holes in the particle arrays, etc. The spatial localization of the plasmon features is generally considered to be delocalized and, when probed by electron beam, decays according to a modified Bessel function of the second kind[15]. Nonetheless, as was alluded to previously, the extent of localization differs by roughly an order of magnitude when comparing single or few particle responses to coupled array modes - that is to say, the predictability is susceptible to the degree of localization and must also be carefully considered.

Based on these considerations, we aim to explore whether local plasmonic properties can be predicted based on the local system geometry. The classical approach to this problem is based on predictive theory, where the plasmonic property of the system is calculated based on approximated materials constants and the latter are refined to maximize theory-experiment matching. Thus, a refined generative model can then be used for the prediction of plasmonic properties in any material configuration and opens a pathway for predictive design of the structures with responses closest to the ones desired. This approach, however, requires well-defined high-veracity models. Factors such as the presence of surface dielectric layers due to specific adsorption, chemisorption, or contamination can invalidate the model and limit its range of applicability (e.g., the model will possess predictive power for specific materials configuration but may fail beyond it, for example, for different surface states).

Here, we explore an alternative approach for the establishing the relationship between local structure and local plasmonic spectra based on machine learning (ML), i.e., establishing the correlative relationship between the particle geometry in the vicinity of the chosen spatial point and plasmonic spectra at the same point. While not having the predictive power of a generative model, such an approach is expected to be valid for in-distribution observations and will thus benefit from larger data sets assuming the absence of observational bias variability (i.e., relevant spectral features depend only on material and location but not on the specific microscope parameters). Compared to generative models, this approach relies only on the observed configurations, and does not necessitate an in-depth knowledge of the material's structure or generative physical behaviors, thus providing a counterpart for classical physical approaches. This methodology can potentially be applied to any arbitrary system to derive structure-property relationships (provided such a relationship exists), both for systems that are too complex to treat theoretically and even in novel materials and geometries where no proven physical model has been established.

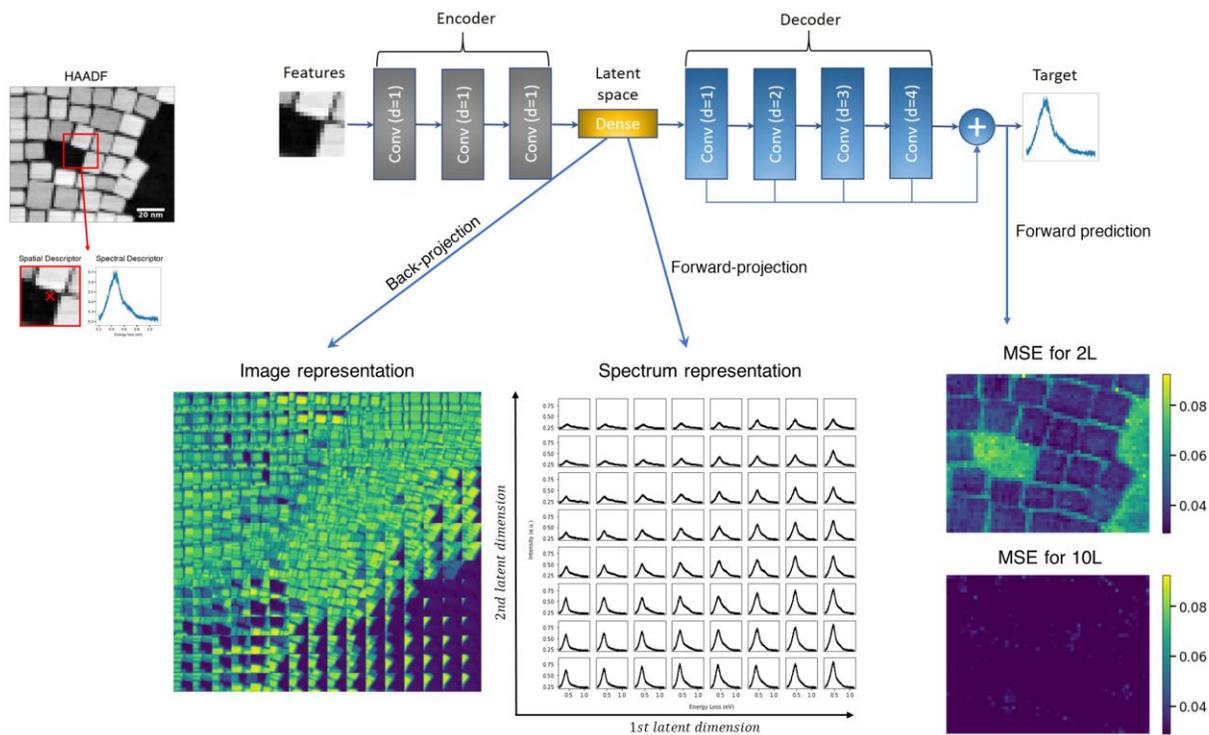

**Figure 3.** The *im2spec* network is trained using the HAADF spatial descriptors and corresponding spectral descriptors. Here, features are sub-images while targets are spectra. The latent space is a compressed information space, which contains features significant within the feature array. When training with two latent dimensions, the latent space can be visualized as a grid in 2D in either an image or spectral representation. 'Conv' denotes a convolutional layer with kernel size of 3 activated by a leaky rectified linear unit with a negative slope of 0.1, 'd' is the dilation rate, and 'Dense' stands for a fully-connected layer whose number of neurons is equal to the number of latent dimensions. Mean squared error (MSE) is shown for using two and ten latent dimensions, 2L and 10L respectively. Note that while errors for two latent variables show clear spatial structure, the error does not exceed 10%. For 10 latent variables, MSE does not exceed 4% and almost spatially uniform, suggesting that the *im2spec* network successfully encodes structure-property relationships. Note that the training for 2L and 10L was performed separately.

The analysis with the *im2spec* network is shown in Figure 3 where the images were parsed into 2714 sub-images with each sub-image containing 16x16 pixels in size. The spectra were used at original resolution, and to isolate plasmon-specific spectra, post-processing was performed to remove the zero-loss peak (ZLP) and the phonon signal arising from the silicon nitride support membrane. The encoder-decoder architecture used here, compressing the images to latent variables, $(image) \rightarrow (latent)$, and then decompressing them into spectra, $(latent) \rightarrow (spectra)$, allows for analysis of the data in the latent space. In classical autoencoders where the feature and target space are the same, this is accomplished by the determining the variability of the data in the latent space, i.e. finding the maximum and minimum values of latent parameters $L_i$, creating the uniformly spaced grid of points in the latent space, and projecting this grid onto the feature space. This representation is particularly convenient for two latent parameters, where the feature evolution in the latent space can be represented as a grid image.

To adapt this approach for the encoder-decoder architectures used here, we note that for the decoder part of the network the same approach can be used. However, a similar analysis can be extended for the encoder part of the network, where we back-project the point in the latent space by finding the feature (or average of several features) closest to the selected point. In other words, to back-project the point $L_1, L_2, .. , L_n$ from the n-dimensional latent space into the original feature space, $(feature) \leftarrow (latent)$, we find images where projection to the latent space is closest to the selected point. Note that in this manner only physically realizable features (or their averages) are visualized as compared to the $(latent) \leftarrow (target)$, projection where the latent point can come from an unpopulated part of the latent space. Similarly, the veracity of back-projection can be ascertained from a comparison of the dispersions (or other measures of variability) in the feature space vs. latent space, i.e., whether closely located points in the latent space back-project to the closest points in the feature space.

The back-projected features for the *im2spec* network for two latent variables for a uniform grid in latent space are illustrated in Figure 3. Note that the adjacent locations in the latent space correspond to similar morphologies of the sub-images, and dissimilar morphologies correspond to well-separated parts of the latent space. For example, the bottom right part of the back-projected image representation in Figure 3 corresponds to different separations from the particle corner, while the top-right region corresponds to different beam positions in the dense part of the film. In this way, the two latent features that were extracted by the network are visualized in a 2D grid. Note that we can specify a different number of latent dimensions when utilizing the autoencoder network, but only the case of two latent features can be represented in a two-dimensional space.

Additional insight into the veracity of the *im2spec* conversion can be derived from analysis of the prediction error. For this, the mean square error between the predicted and actual spectra is calculated and plotted in the original image plane. In this analysis, of interest is the absolute value of the mean square error (MSE), and spatial localization of the MSE signal. The data in Fig. 3 suggest that for two latent variables the prediction error can be on the of order of 10%, but clearly shows the spatial structures associated with the edges. This behavior illustrates that two latent variables are insufficient to encode all the correlative relationships in this system, and that there is a difference between the physical mechanisms operating at the edges and in the volume of the film. In comparison, the error surface for 10 latent variables shows errors on the order of only several percent with almost uniform spatial distribution, suggesting that this encoding is sufficient to build a correlative model universally applicable for these data.

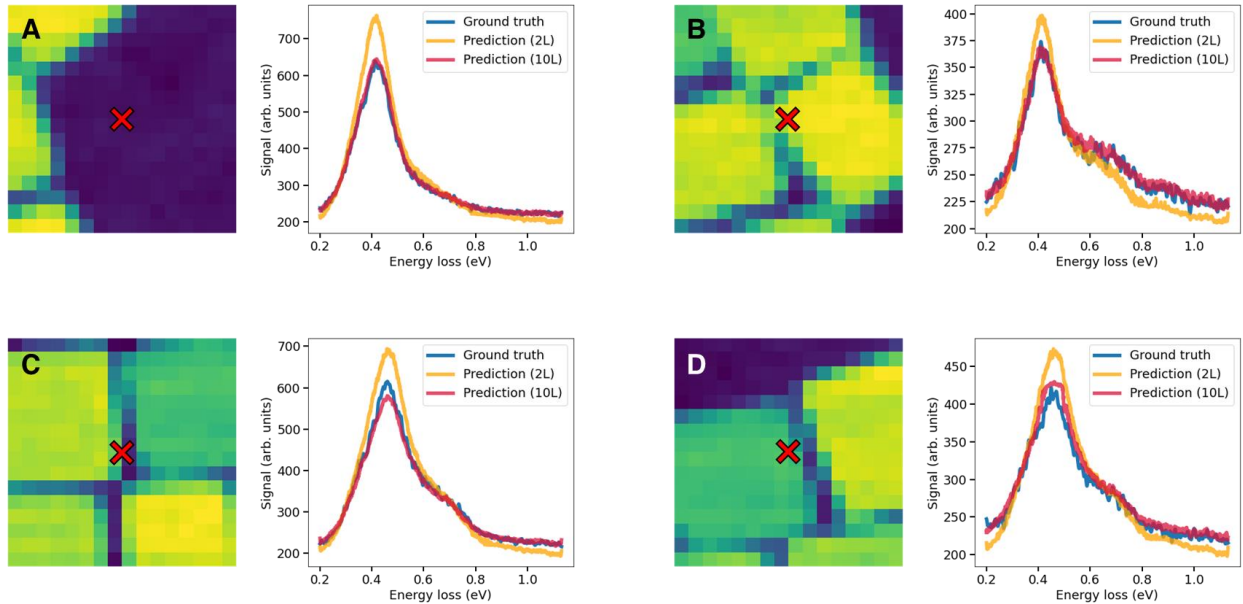

**Figure 4.** *Im2spec* network applied to cubic plasmon array system. Red crosshair indicates from where the spectra are taken. (a) Depicts behavior in a void, (b) on a particle, (c) in an interparticle gap, and (d) in an interparticle gap near an edge. Note that (b) is only case in which the highest energy feature near 0.9 eV is observable since this is a bulk plasmon mode. The *im2spec* network accurately predicts this weak bulk feature when the number of latent dimensions is sufficient.

Finally, a comparison of the predicted and observed spectra for several locations are shown in Figure 4. Here, we highlight regions from the SI showing different aspects of nanoscale disorder: one region from a hole in the array (a), one located on a particle within the array (b), one from regular part of the array in an interparticle gap (c), and one located in an interparticle gap near an edge (d). For each region, we show the sub-image used as the input into the im2spec network, as well as the spectra both taken directly from the central pixel of the subregion in the SI, as indicated by the red crosshair, as well as the *im2spec* predictions trained with both 2 and 10 latent space variables (2L and 10L, respectively). We note the network has never encountered these particular sub-images previously, as it was trained on a different portion of the data cube. In all circumstances, the 10L *im2spec* network both qualitative and quantitatively reproduces the ground-truth spectrum solely from the local configuration of particles, demonstrating the capability of directly retrieving structure-property relationships from this data-driven approach. The importance of using a high-dimensional latent space is also observed as there are serious quantitative discrepancies that appear in the 2L network that are not observed in the 10 L network.

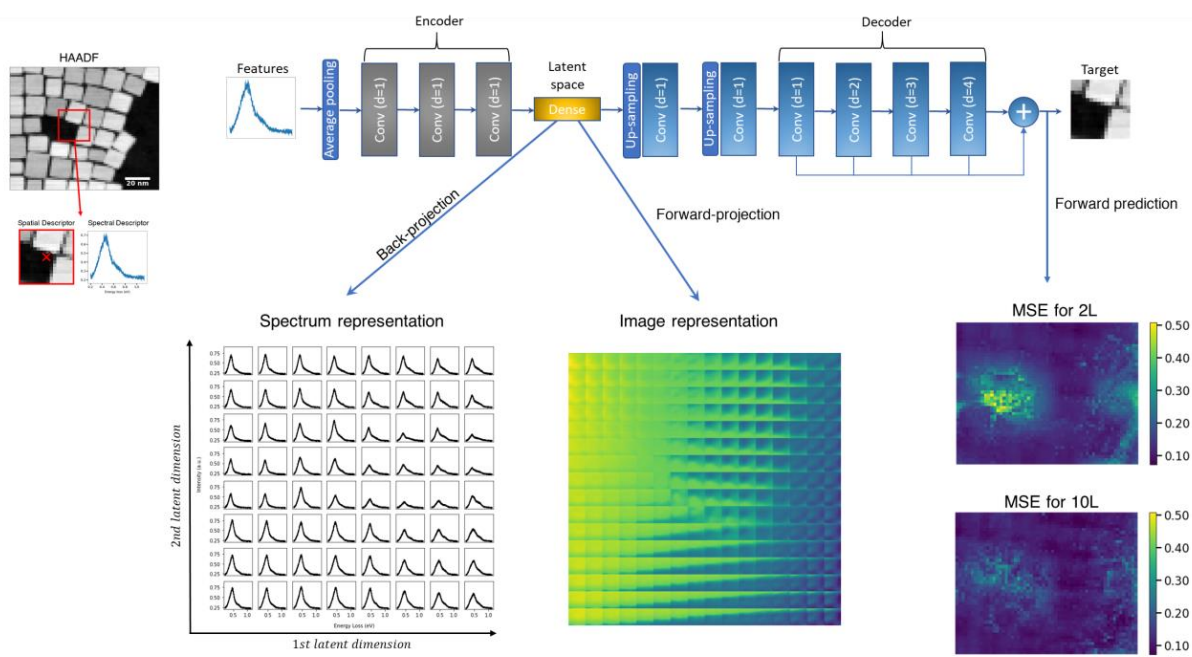

**Figure 5.** The *spec2im* network is trained in a similar way to *im2spec*, also using the HAADF spatial descriptors and corresponding spectral descriptors, but with features and targets exchanged. The latent space is a compressed information space, which contains features significant within the feature array. When training with two latent dimensions, the latent space can be visualized as a grid in 2D in either an image or spectral representation. 'Conv' denotes a convolutional layer with kernel size of 3 activated by a leaky rectified linear unit with a negative slope of 0.1, 'd' is the dilation rate, and 'Dense' stands for a fully-connected layer whose number of neurons is equal to the number of latent dimensions. Mean squared error (MSE) is shown for using two and ten latent dimensions, 2L and 10L respectively.

A similar approach can be explored to establish the relationship between the spectra and images, as realized in the *spec2im* network. The structure of the *spec2im* is similar to that of *im2spec* as shown in Figure 3, with the interchanged encoder and decoder architectures. The latent space representations corresponding to a uniform grid in latent space projected to image space, $(latent) \rightarrow (image)$, is shown in Figure 5. Note that the latent space clearly clusters possible particle configurations, by virtue of the fact that individual tiles in the latent spaces show a variety of particles positions which exist in similar regions of the 2D space. In fact, detailed analysis of the reconstructed images across the latent space clearly illustrates that the autoencoder disentangles particle shape representations, with consistent and smooth changes of particle shapes across selected dimensions.

The structure of the spatial error maps is shown in Figure 5. Note that while in this case we generally expect high prediction errors (due to the rotation uncertainty), the reconstruction is surprisingly good. The average reconstruction error is 15.8% for 2D latent space and 12.7% for 10D latent space. The errors are higher in the vicinity of the edges, as expected. For a lower resolution image where the sub-image size is larger than effective particle size, the quality of prediction is considerably higher since the latent variables can encode relevant details, as shown in Figure S3. Calculation of the cross correlation coefficient and the MSE between the structural HAADF-STEM image and the error maps in Figures 3 and 5 reveals that with increasing number of latent dimensions, generally more features are matched in the error maps to their corresponding structural HAADF image, as is indicated by both a decreased MSE and an increased and more positive correlation coefficient which is observed in Figure S1. This indicates that there are regions where the correlative relationship between structure and plasmonic properties is better defined. The higher dimensional latent space therefore encodes these features better and we are able to assess an appropriate number of latent dimensions from the preceding analysis. These regions are potentially of physical exploratory interest.

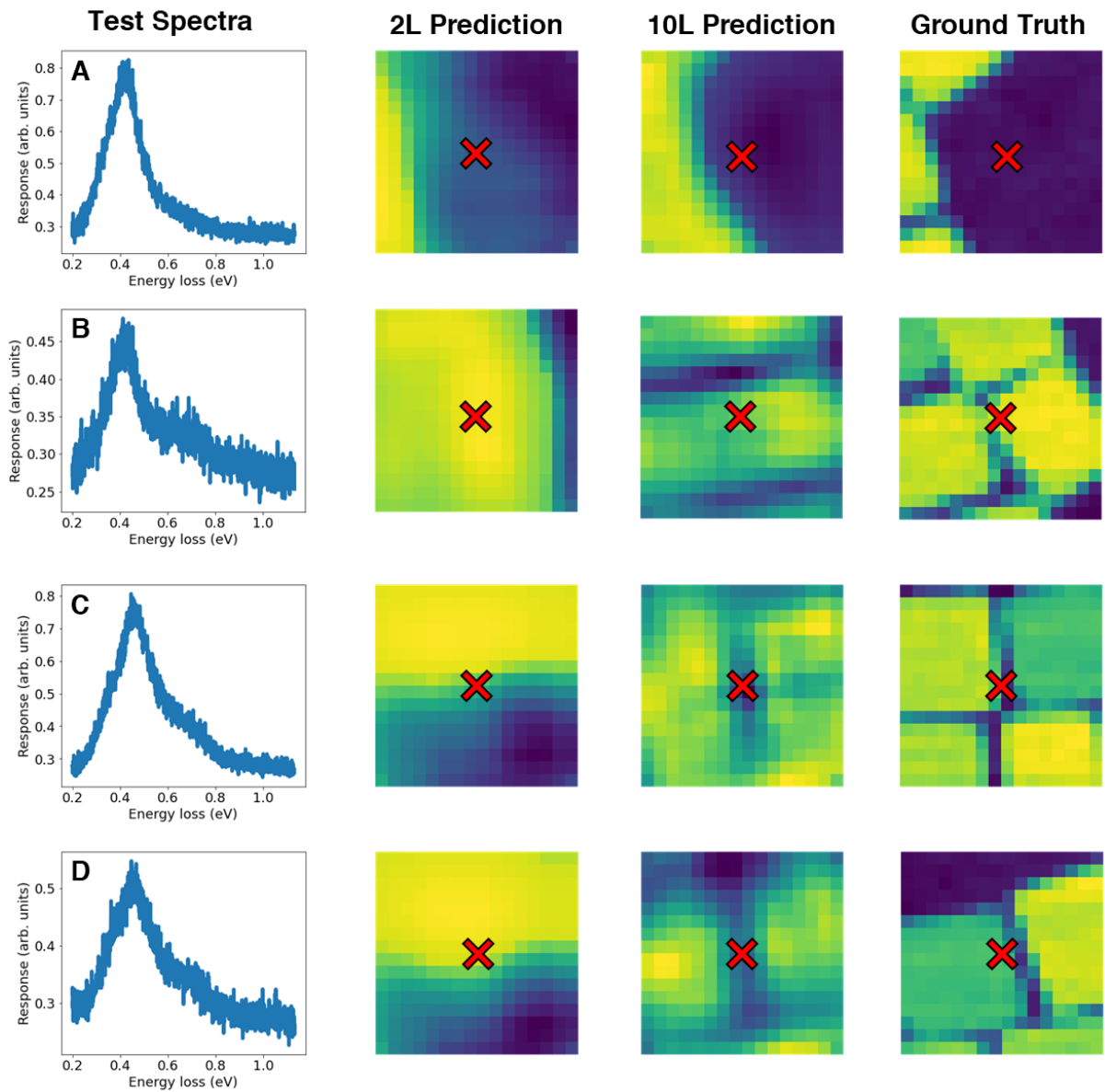

**Figure 6.** Predictions of the local images from spectra trained using two and ten latent dimensions (a-d) illustrating slightly different behaviors – voids, gaps, and edges. These structure-property relationships are evidently encoded in the high dimensional latent space. Note the test spectra are not arbitrary – they have a direct correspondence to the real system (ground truth sub-images) but have never been directly encountered by the network.

Several examples of image predictions from spectra are shown in Figure 6 (a-d). Note that while the exact morphology cannot be reproduced, general morphology including larger scale heterogeneity such as the edge of the array and the hole in the array is reproduced remarkably well.

Not only does the *spec2im* network predict the existence of these features, it unexpectedly and rather astonishingly predicts in several cases in the same rotational orientation and areal fraction as in the ground truth sub-image. However, one possible explanation for this is a slight specimen tilt, which may allow different plasmon features to be detected that depend on rotation. The ability to procure such features in the case of the inherently delocalized plasmon should not be understated – it is remarkable that general correlative behaviors can be learned from local sampling. This strongly establishes the ability of the encoder-decoder networks to generate reliable structure-property relationships from highly localized sampling.

To summarize, we illustrate a ML-based workflow to establish the correlative relationships between nanoparticle configurations in the proximity of a chosen spatial location and the associated plasmonic spectra. This approach is complementary to that based on generative modeling. While the model requires detailed knowledge of the system composition to generalize for unobserved configurations, the proposed *im2spec* and *spec2im* networks allow for predictions of samples drawn from similar distributions and can generalize the structure property relationships without prior knowledge or observational biases. We expect the features learned in the latent spaces to be transferrable between data sets assuming observational conditions are the same, i.e., train network with one data set, and predict using a new one. Of course, constraints surround this expectation, namely, particle size and concentrations should not deviate far from those in the data set used in training.

With careful experimentation, the derived correlative relationships can be treated as universal, and used as a basis for the refinement of the generative theoretical models, for example via Bayesian optimization in the latent space or simply generating the structure-spectrum pairs. For non-universal cases (i.e. dependent on microscope setting, indicative of the presence of observational biases or latent variables), a similar analysis can be used to establish variability and possible behaviors and anomalies within the image (or set of images obtained under similar conditions), allowing for qualitative or at least semiqualitative explorations of relevant physical behaviors comparable to standard unmixing methodologies which also do not account for phenomena outside of the region of interest. In both cases, the knowledge of possible structure-property relationships allows to establish the range of possible responses within a given materials system. Also, regions with large deviations indicate the presence of something unusual or exciting – likely worth pursuing further.

Perhaps more importantly, fundamentally new opportunities can be presented for cases where the particles can be manipulated either globally (e.g., via chemical functionalization) or especially locally (e.g., *ex-situ* via scanning probe microscopy, or *in-situ* via an electron beam). In this case, the knowledge of correlative structure property relationships from *im2spec* and *spec2im* networks will allow the creation of structures with desired properties. These approaches can include several iterative steps when we learn laws, make structures, then refine those laws to design what we want.

**Acknowledgements:**

This effort (ML and STEM) is based upon work supported by the U.S. Department of Energy (DOE), Office of Science, Basic Energy Sciences (BES), Materials Sciences and Engineering Division (K.M.R., S.V.K.) and was performed and partially supported (J.A.H., M.Z.) at the Oak Ridge National Laboratory's Center for Nanophase Materials Sciences (CNMS), a U.S. Department of Energy, Office of Science User Facility. S.H.C. acknowledges NSF grant CHE-1905263. We thank Andrew Lupini for his fruitful discussions and valuable advice in the manuscript.

**Author contributions**: (K.M.R.) acquired and analyzed experimental STEM-EELS data using the architectures discussed in the text, wrote and edited the manuscript, (S.H.C.) synthesized and prepared the sample, (S.V.K.) conceived and led the project, and wrote part of the manuscript, (J.A.H.) assisted in data acquisition and contributed significantly to the manuscript, (M.Z.) developed the machine learning algorithms and processes for use with STEM-EELS data, and helped write the manuscript.

**Competing interests**: The authors declare no competing interests.

**Data and materials availability**: The codes used in this manuscript can be found at: https://git.io/JURbN

## Supplementary Materials

**Materials and Methods**

*Imaging and spectroscopy*: A monochromated aberration-corrected NION microscope was operated at an accelerating voltage of 60 kV, with a probe current on the order of 20 pA, and a convergence angle of 30 mrad. The full width half maximum of the zero loss peak (ZLP) after monochromation was ~40 meV, with the potential to be reduced to as low as 5 meV, however this was kept higher to maintain a high signal to noise ratio. Pixel dwell times for spectrum images were 100 ms. The sample pressure is kept at a stable $10^{-9}$ Torr and is especially important for longer dwell times to minimize contamination of the sample, which can have a significant impact on the plasmon response. Post processing to remove the zero-loss peak and the phonon signal arising from the silicon nitride support membrane was performed by fitting a two-term exponential power law to the zero-loss peak, and truncating the energy axis below 200 meV, respectively.

A*utoencoder network details*: The *im2spec* and *spec2im* networks are implemented here via the PyTorch deep learning library. The features are passed through multiple convolutional layers and a fully connected ("dense") layer to create a latent representation, which is then used to generate a target output via a spatial pyramid of dilated convolutions. We note that using a simpler decoder network (e.g. with the same structure as the encoder) also works but results in a lower prediction accuracy. The *spec2im* has a similar structure. Compared to *im2spec*, the difference is that we use the average pooling prior to the convolutional block to reduce the length of input vector by a factor of 16 and add two up-sampling layers before the dilated block in the decoder to help with the decompression of latent features. The mean squared error loss was optimized using the Adam optimization technique[22] with a learning rate of 0.001 for both neural networks.

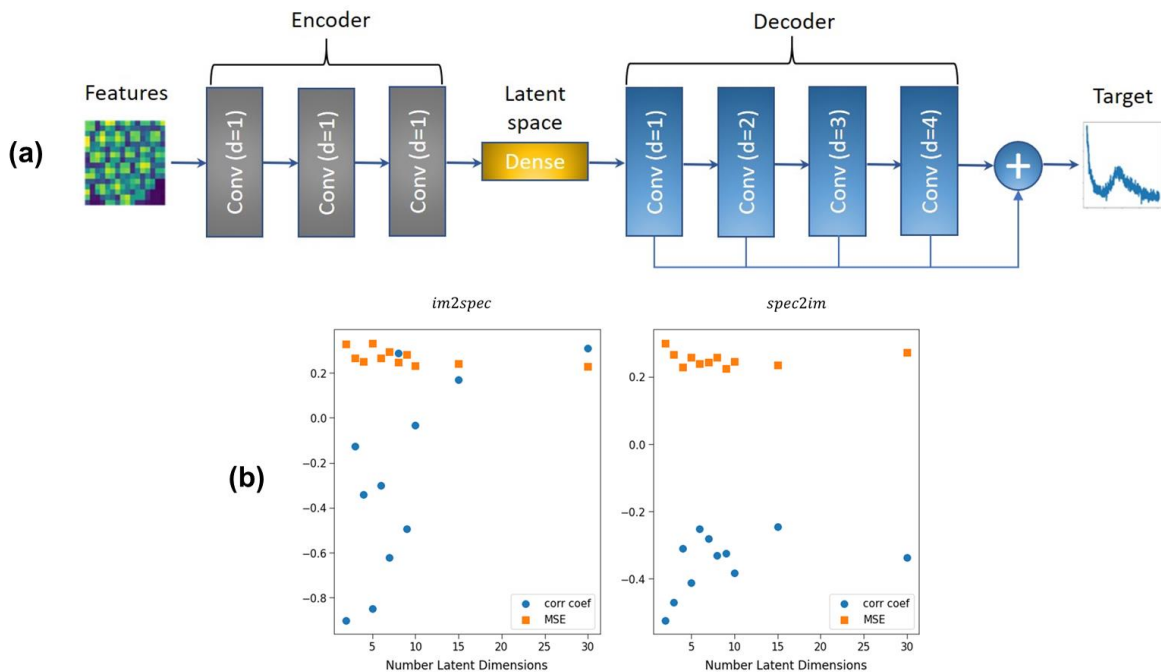

**Figure S1.** (a) Schematics of *im2spec* neural network. 'Conv' denotes a convolutional layer with kernel size of 3 activated by a leaky rectified linear unit with a negative slope of 0.1, 'd' is the dilation rate, and 'Dense' stands for a fully-connected layer whose number of neurons is equal to the number of latent dimensions. (b) Effect of number of latent dimensions on the predictability by calculating correlation coefficient and mean squared errors (MSE) between error maps of the higher resolution and their corresponding structural ADF image. The correlation coefficient trends toward an improved correlation, as well as a decreased MSE, with an increase in number of latent dimensions for both *im2spec* and *spec2im* networks.

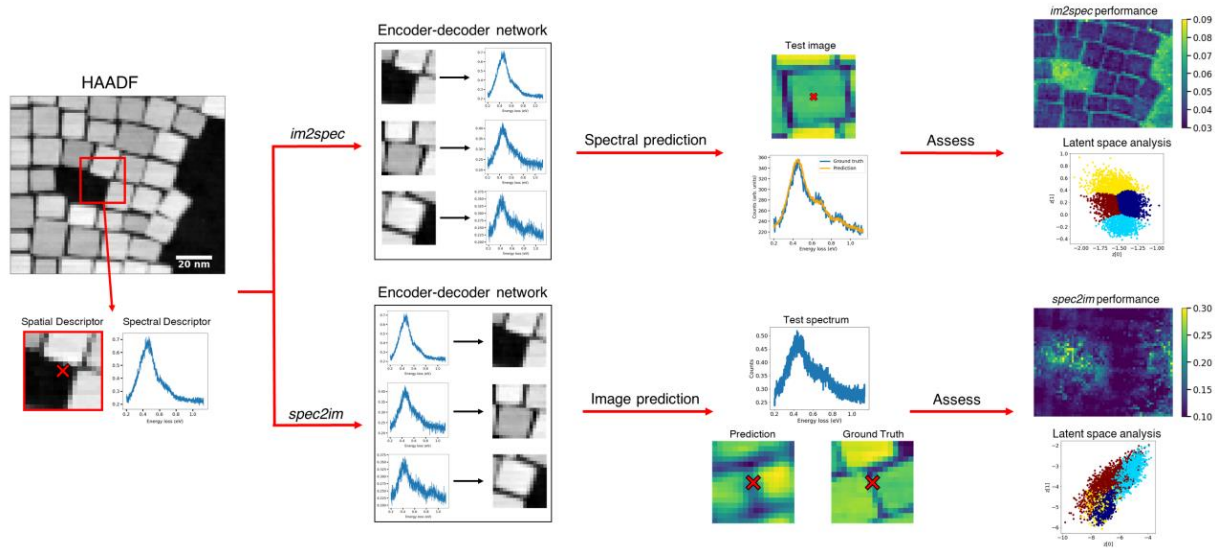

**Figure S2**. Full network architecture, showing both *im2spec* and *spec2im*. Assessment is performed by MSE in real space and will depend on number of latent dimensions chosen. The latent space itself can be analyzed by clustering methods to highlight dominant features.

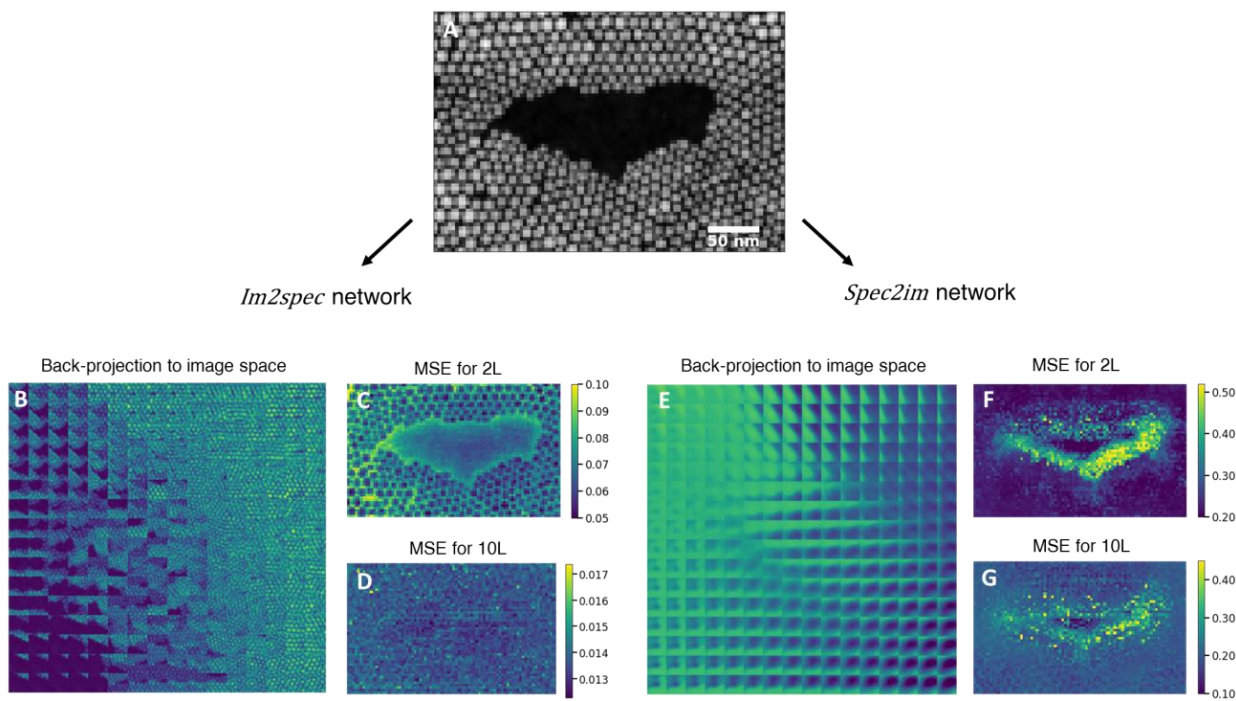

**Figure S3**. Network performance on different data set acquired at a lower number of pixels per particle. (a) HAADF-STEM image; results for *im2spec* network shown in (b-d); (e-g) spec2im

network results. (b,e) Show back-projection from latent space into an image representation, while (c,d,f,g) portray error maps for specified number of latent dimensions.